\documentclass[aps,prd,preprint,superscriptaddress]{revtex4}

\usepackage{latexsym}
\usepackage{amsmath,amssymb}
\usepackage{graphicx}
\usepackage{subfigure}

\begin{document}


\title{Gauged Lifshitz scalar field theories in two dimensions}

\author{Myungseok Eune}
\email[]{younms@sogang.ac.kr}
\affiliation{Research Institute for Basic Science, Sogang University, Seoul, 121-742, Korea}

\author{Wontae Kim}
\email[]{wtkim@sogang.ac.kr}
\affiliation{Center for Quantum Spacetime, Sogang University, Seoul 121-742, Korea}
\affiliation{Department of Physics, Sogang University, Seoul 121-742, Korea}
\affiliation{School of Physics, Korea Institute for Advanced Study, Seoul 130-722, Korea}

\author{Edwin J. Son}
\email[]{eddy@sogang.ac.kr}
\affiliation{Center for Quantum Spacetime, Sogang University, Seoul 121-742, Korea}

\date{\today}

\begin{abstract}
  We present two-dimensional gauged Lifshitz scalar field theories by
  considering the duality relation between the source current and the
  Noether current.  Requiring the duality partially, we obtain a
  gauged model which recovers the bosonized Schwinger model for the IR
  limit. For the exact duality, however, the source current is not
  conserved, which means that the resulting theory is anomalous, so
  that the number of degrees of freedom is increased. The second model is
  consistently formulated by adding the Wess-Zumino type action to
  maintain the gauge invariance.
\end{abstract}


\maketitle

\section{Introduction}
\label{sec:intro}

A Lifshitz scalar field theory~\cite{lifshitz,hls,Horava:2008jf}
has been studied in the condensed matter physics as a description of
tricritical phenomena involving spatially modulated phases.  The
Lifshitz index $z$ reflects the anisotropic scaling between space and
time, $x\to bx$ and $t\to b^z t$, and the Lifshitz scalar theory
describes a free field fixed point $z=2$ in four dimensions.  Adding a
relevant deformation can make the theory flow to a fixed point $z=1$,
which is in fact satisfied by the relativistic theory so that the
Lorentz invariance emerging as an accidental symmetry at long
distances.  Recently, a renormalizable theory of gravity has been
suggested by Ho\v{r}ava, as called Ho\v{r}ava-Lifshitz
gravity~\cite{Horava:2008ih,Horava:2009uw}. Due to the
difference between the scaling dimensions of space and time, the
Lagrangian is composed of usual second derivatives with respect to time and
higher-order derivatives with respect to space. Of course, in
the UV region, the explicit Lorentz covariance is broken.

On the other hand, a two-dimensional Dirac fermion coupled to the
gauge field can be completely bosonized~\cite{Coleman:1974bu,
  Mandelstam:1975hb, Halpern:1975nm, Halpern:1975jc}, and thus
Schwinger model(SM) can be remarkably described by a scalar field
coupled to a gauge field, which gives a single massive
mode~\cite{Coleman:1974bu, Mandelstam:1975hb, Halpern:1975nm,
  Halpern:1975jc,Treiman:1986ep, Boyanovsky:1987ad}.
By the way, from the beginning, it is possible to realize this
model in the bosonic regime. First, from the Noether current
associated with the global symmetry, one can define the gauge
invariant sector of the Noether current, though it is no longer
conserved due to the axial anomaly when we consider the gauge
field coupling. Then, the dual current through the duality
relation of $J_{\rm V}^\mu =\epsilon^{\mu\nu} J^{\rm A}_\nu$ turns
out to be nothing but the conserved gauge invariant vector current
which can couple to the gauge field, where $J_{\rm V}^\mu$ and
$J^{\rm A}_\nu$ are the gauge invariant conserved vector current
and the gauge invariant non-conserved axial current, respectively,
and $\epsilon_{01} = +1$. The vector current as a source current
can couple to the gauge field, eventually, which yields the SM
\cite{Treiman:1986ep}.

Now, one considers a single Lifshitz scalar field which recovers
the Lorentz covariance for the IR limit, then it is natural to
extend this free Lifshitz field theory to the gauged Lifshitz
scalar field theory which gives the SM for the IR limit.  In this
paper, we would like to construct a gauged Lifshitz scalar field
theories in two dimensions adopting the above-mentioned method.
For this purpose, we define Noether and source currents for
convenience, where they become the genuine axial and vector
currents in QED for some limits. We first find the gauge invariant
Noether current corresponding to the axial current in
section~\ref{sec:gaugedLifshitz}. Then, the conserved source
current can be obtained by imposing the duality relation
partially. Then, we shall obtain the gauged Lifshitz scalar field
theory which is continuously interpolated with the SM along with
some conditions for the boundedness of the Hamiltonian. For the IR
limit, we can recover the well-known SM. In
section~\ref{sec:dual}, we study a Lifshitz scalar field coupled
to the source current satisfying the full duality of $J_{\rm
S}^\mu = \epsilon^{\mu\nu} J^{\rm N}_\nu$, where $J_{\rm V}^\mu =
\epsilon^{\mu\nu} J^{\rm A}_\nu$ for the IR limit. But the source
current is not conserved, so that the theory becomes anomalous.
Moreover, there does not exist a bounded Hamiltonian, which breaks
unitarity. So, we introduce an additional constraint to resolve
the unitarity problem, and add the Wess-Zumino(WZ) action to
recover the local gauge invariance~\cite{Wess:1971yu,
Faddeev:1986pc, Harada:1986wb, Falck:1986rt,
  Harada:1989qp, Kye:1991uu}.  The resulting theory is consistent, but
the model is different from the case of the partial duality.  Finally,
summary is given in section~\ref{sec:conclusion}.

\section{Gauged Lifshitz scalar with the partial duality}
\label{sec:gaugedLifshitz}
The action describing a Lifshitz scalar
$\phi$ up to the fourth derivatives is defined by
\begin{equation}
  \label{S:L}
  S_{\rm L} = \frac12 \int d^2 x \left( \partial_\mu \phi \partial^\mu
    \phi + \beta \phi''^2 \right),
\end{equation}
where $x^\mu = (t,x)$, and the Minkowski metric is $\eta_{00} =
1$. The parameter $\beta$ is an arbitrary constant, which can be
fixed based on physical requirements in later. The overdot and the
prime denote the derivatives with respect to $t$ and $x$,
respectively. Introducing an auxiliary field $\lambda$ for
conveniences, the action (\ref{S:L}) can be written as
\begin{equation}
  \label{S:0}
  S_{\rm L} = \frac12 \int d^2 x \left[ \partial_\mu \phi \partial^\mu
    \phi + \beta (\lambda^2 - 2 \lambda' \phi') \right].
\end{equation}
Now, the equations of motion are obtained as
\begin{align}
  \lambda &= - \phi'', \label{eom:scalar:lambda} \\
  \Box \phi &=  \beta \lambda'',
\end{align}
where $\Box = \partial_\mu \partial^\mu$. Since the action~(\ref{S:0})
is invariant under the global transformation given by $\delta \phi =
\mathrm{const.}$ and $\delta \lambda = 0$, the Noether current can be
calculated as
\begin{equation}
  \label{J:A}
  J_{\rm N}^\mu = \partial^\mu \phi - \beta \delta^\mu_1 \lambda',
\end{equation}
which is so-called the axial current in the bosonized SM for
$\beta=0$~\cite{Coleman:1974bu, Mandelstam:1975hb, Halpern:1975nm,
  Halpern:1975jc}. Similarly to the SM, let us define the source
current which couples to the gauge field, then it may be defined by
the duality relation between the source current and the Noether
current as
\begin{equation}
  J_{\rm S}^\mu = \epsilon^{\mu\nu} J^{\rm N}_\nu.
\end{equation}
The reason why we want to define the source current in terms of the
duality relation is that it should be coincident with the vector
current in the bosonized SM for $\beta\to0$, for instance, $J_{\rm
  V}^\mu = \epsilon^{\mu\nu} J^{\rm A}_\nu$.  However, it breaks the
conservation law as seen from $\partial_\mu J_{\rm S}^\mu = - \beta
\dot\lambda' \ne 0$.

In what follows, the source current satisfying the conservation law
can be defined by requiring the dual relation partially between the
source and the Noether currents; type I ($J_{\rm S}^0 = \epsilon^{01}
J^{\rm N}_1$ but $J_{\rm S}^1 \ne \epsilon^{10} J^{\rm N}_0$) and type
II ($J_{\rm S}^0 \ne \epsilon^{01} J^{\rm N}_1$ but $J_{\rm S}^1 =
\epsilon^{10} J^{\rm N}_0$). Then, the source current can be written
as
\begin{equation}
  \label{J:v:conserved}
  J_{\rm S}^\mu = \epsilon^{\mu\nu} \partial_\nu ( \phi + b \lambda),
\end{equation}
where $\partial_\mu J_{\rm S}^\mu =0$. The constant, $b$ should be
chosen as $\beta$ for the type I or $0$ for the type II. Then, we can
write down the action:
\begin{align}
  S_\text{(partial)} &= S_{\rm L} +\int d^2 x \left( -e J_{\rm S}^\mu A_\mu
  \right) +
  S_\text{Max}, \label{S:conserved}
\end{align}
where $S_\text{Max} = \int d^2 x \left( - \frac14 F_{\mu\nu}
  F^{\mu\nu} \right)$, and $A_\mu$ and $F_{\mu\nu}$ are the gauge
field and the field strength, respectively.
As a comment, the Noether current in the presence of the
gauge field is modified as
\begin{equation}
  j^\mu = J_{\rm N}^\mu + e \epsilon^{\mu\nu} A_\nu,
\end{equation}
which satisfies $\partial_\mu j^\mu = 0$. However, this Noether
current is not invariant under the local gauge transformation
given by $\delta A_\mu = \frac{1}{e} \partial_\mu \Lambda$, so
that the gauge invariant current should been chosen as $ J_{\rm
N}^\mu$ in Eq.~(\ref{J:A}). This gauge invariant current has an
axial anomaly, \textit{i.e.,} $\partial_\mu J_{\rm N}^\mu = - e\,
\epsilon^{\mu\nu} \partial_\mu A_\nu$~\cite{Treiman:1986ep}.

The equations of motion from the action~(\ref{S:conserved}) are
obtained as
\begin{align}
  &\Box \phi - \beta \lambda'' + e
  \epsilon^{\mu\nu} \partial_\mu A_\nu = 0, \label{eom:conserved:phi} \\
  &\beta (\lambda+\phi'') - be \epsilon^{\mu\nu} \partial_\mu A_\nu =
  0, \label{eom:conserved:lambda} \\
  &\partial_\mu F^{\mu\nu} + e \epsilon^{\mu\nu} \partial_\mu (
  \phi + b \lambda) = 0. \label{eom:conserved:A}
\end{align}
By eliminating the scalar field and the auxiliary field from
Eqs.~(\ref{eom:conserved:phi}), (\ref{eom:conserved:lambda}), and
(\ref{eom:conserved:A}), we get
\begin{equation}
  \label{eom:conserved:eff}
  \left[ \Box + \beta \partial_x^4 +
    \frac{e^2 (1 - b \partial_x^2)^2}{1-be^2} \right] {}^*F = 0, 
\end{equation}
where ${}^*F = \frac12 \epsilon^{\mu\nu} F_{\mu\nu}$, so that there
exists only one non-tachyonic massive mode with mass
$m^2=e^2/(1-be^2)$ for $b<1/e^2$. Note that the mass of the physical
field for the type II ($b=0$) is compatible with that of the SM. As
for the type I ($b = \beta$), the higher-derivative correction gives
rise to higher massive state than that of the SM.

In order to examine the boundedness of the Hamiltonian, we perform
the constraint analysis in terms of the Hamiltonian formulation.
The conjugate momenta with respect to $A_0$, $A_1$, $\phi$, and
$\lambda$ are given by
\begin{align}
  \Pi^0 &= 0, \qquad
  \Pi^1 = \dot{A_1} - A'_0, \\ 
  \Pi_\phi &= \dot{\phi} - e A_1, \qquad
  \Pi_\lambda = -b eA_1,
\end{align}
respectively, from which we read the primary constraints as
follows \cite{Dirac:1964LQM}:
\begin{align}
  \Omega_1 &\equiv \Pi_0 \approx 0, \quad
  \Omega_2 \equiv \Pi_\lambda + be A_1 \approx 0. \label{conserved:Omega:1,2}
\end{align}
Then, the primary Hamiltonian is obtained as
\begin{equation}
  H_p = \int dx (\mathcal{H}_c + \eta_1 \Omega_1 + \eta_2 \Omega_2),
\end{equation}
where the canonical Hamiltonian density is given by
\begin{align}
  \mathcal{H}_c = &\frac12 (\Pi_\phi + e A_1)^2 + \frac12 \phi'^2 -
  \frac12 \beta( \lambda^2 - 2\lambda' \phi') -e ( \phi' + b \lambda')
  A_0 + \frac12 (\Pi^1)^2 + \Pi^1 A'_0,
\end{align}
and $\eta_i$'s are the multiplier fields. From the time evolution of the
primary constraints (\ref{conserved:Omega:1,2}), we obtain the
secondary constraints:
\begin{align}
  \Omega_3 &\equiv (\Pi^1)' + e(\phi' + b \lambda') \approx
  0, \label{conserved:Omega:3} \\
  \Omega_4 &\equiv \beta(\lambda + \phi'') + be \Pi^1 \approx 0. \label{conserved:Omega:4}
\end{align}
Since the constraint algebra is first class which reflects the local
gauge symmetry, we can choose the Coulomb gauge:
\begin{equation}
 \Omega_5 \equiv A'_1 \approx 0. \label{conserved:Omega:5}
\end{equation}
Through the time evolution of this gauge fixing
condition~(\ref{conserved:Omega:5}), we obtain one more constraint:
\begin{equation}
  \label{conserved:Omega:6}
  \Omega_6 \equiv (\Pi^1)' + A''_0 \approx 0.
\end{equation}
Consequently, the net degrees of freedom is two in the phase space,
and the reduced Hamiltonian at the constraint surfaces is obtained as
\begin{equation}
  H_\text{red} = \frac12 \int dx \left( \Pi_\phi^2 +
    \frac{e^2}{1-be^2} \phi^2 + \frac{1 + be^2}{1-be^2} \phi'^2 +
    \frac{\beta}{1-be^2} \phi''^2 \right),
\end{equation}
which is positive for $0 < \beta < 1/e^2$ for the type I ($b = \beta$)
and $\beta>0$ for the type II ($b=0$).


As a result, the gauged Lifshitz field theory can be
obtained from the coupling to the conserved source current
satisfying the partial duality, which gives the well-defined SM
for $\beta\to0$. Next, we are going to study the gauge coupling to the
source current defined by the exact duality.

\section{Gauged Lifshitz scalar with the exact duality}
\label{sec:dual}

We consider a Lifshitz scalar coupled to the gauge field by
requiring the exact duality, \textit{i.e.,} $J_{\rm S}^\mu =
\epsilon^{\mu\nu} J^{\rm N}_\nu$ with the Noether current given by
Eq.~(\ref{J:A}), which gives $J_{\rm S}^\mu = \epsilon^{\mu\nu} (
\partial_\nu \phi - \eta_{\nu 1} \beta \lambda')$. As mentioned
earlier, the source current is not conserved, which implies that
the action is not gauge invariant. So, we have to add the WZ
action to cancel the gauge noninvariace~\cite{Wess:1971yu,
Faddeev:1986pc, Harada:1986wb,
  Falck:1986rt, Harada:1989qp, Kye:1991uu}. Then, the total action is
assumed to be
\begin{align}
  \tilde{S}_\text{(exact)} &= S_{\rm L} + \int d^2 x \left( -e J_{\rm S}^\mu A_\mu
  \right) + S_\text{Max} + \tilde{S}_\text{WZ}, \label{S:dual}
\end{align}
where \begin{equation}
  S_\text{WZ} = \int d^2 x\, \beta \lambda' \dot{\theta}, \label{S:WZ}
 \end{equation}
and $\theta$ is a scalar field. Then, the total action is
invariant under the local gauge transformation implemented by
$\delta A_\mu = \frac{1}{e} \partial_\mu \Lambda$ and $\delta
\theta = - \Lambda$.

The Noether current in the presence of the gauge field given by $
j^\mu = J_{\rm N}^\mu + e \epsilon^{\mu\nu} A_\nu $ is not gauge
invariant, and the gauge invariant sector of the Noether current
of Eq.~(\ref{J:A}) is still anomalous, $\partial_\mu J_{\rm N}^\mu
= - e \epsilon^{\mu\nu} \partial_\mu A_\nu$. On the other hand,
the source current with the help of the equation of motion from
the WZ action is automatically conserved, $\partial_\mu J_{\rm
S}^\mu = 0$.

In order to get the reduced Hamiltonian and study the boundedness
of it, let us first obtain the conjugate momenta as follows:
\begin{align}
  \Pi^0 &= 0, \qquad
  \Pi^1 = \dot{A}_1 - A'_0, \\
  \Pi_\phi &= \dot{\phi} - eA_1, \quad
  \Pi_\lambda =0, \quad
  \Pi_\theta = \beta \lambda',
\end{align}
which yields three primary constraints:
\begin{align}
  \Omega_1 \equiv \Pi^0 \approx 0, \quad
  \Omega_2 \equiv \Pi_\lambda \approx 0, \quad
  \Omega_3 \equiv \Pi_\theta - \beta \lambda'
  \approx 0.
\end{align}
Then, the primary Hamiltonian in terms of the Legendre
transformation is written as
\begin{equation}
  H_{p} = \int dx \left(\mathcal{H}_{c} + \sum_{i=1}^3 \eta_i \Omega_i
    \right),
\end{equation}
where the canonical Hamiltonian is given by
\begin{equation}
  \mathcal{H}_{c} = \frac12 [(\Pi_\phi + eA_1)^2 + \phi'^2
  - \beta(\lambda^2 - 2\lambda'\phi')] - e(\phi' + \beta
  \lambda') A_0 + \frac12 (\Pi^1)^2 + \Pi^1 A'_0.
\end{equation}
Now, the gauge fixing condition is taken as
\begin{equation}
  \Omega_4 \equiv \theta \approx 0.
\end{equation}
Then, from the time evolution of the primary constraints and the gauge
fixing condition, we obtain the secondary constraints as
\begin{align}
  \Omega_5 &\equiv (\Pi^1)' + e( \phi' + \beta \lambda') \approx
  0, \\
  \Omega_6 &\equiv \lambda + \phi'' - eA'_0 \approx 0.
\end{align}
By applying all constraints to the primary Hamiltonian, the
reduced Hamiltonian can be obtained as
\begin{equation}
  \label{H:red:dual}
  H_{\rm red} = \frac12 \int dx \left\{ (\Pi_\phi + eA_1)^2 +
    \phi'^2 + \beta (\phi''^2 - e^2 A'^2_0) + e^2 \left[
      \phi - \beta (\phi'' - e A'_0) \right]^2 \right\},
\end{equation}
which is not positive definite unless $\beta \ne 0$ even in spite
of the gauge invariance of the action. It means that the gauged
Lifshitz theory satisfying the exact duality relation between the
source current and the Noether current is not unitary. So, we need
more elaborations to obtain a positive definite Hamiltonian in
this case.

From the reduced Hamiltonian~\eqref{H:red:dual}, one can find a
constraint to make the Hamiltonian to be positive. If we take $
\phi'' + e A'_0=0$, then the reduced Hamiltonian can be positive
definite, however, it is still problematic since the gauge
invariance of the total action is lost. To implement the
constraint gauge invariant fashion, the WZ action to recover the
local gauge symmetry should be modified. To implement the new
constraint consistently, we write the new total action as
\begin{align}
  S_\text{(exact)} &= S_{\rm L} + \int d^2 x \left( -e J_{\rm S}^\mu A_\mu
  \right) + S_\text{Max} + S_\text{cons} + S_\text{WZ}, \label{S:U} \\
  S_\text{cons} &= \int d^2 x \, \beta \xi' (\phi' + e A_0), \label{S:cons} \\
  S_\text{WZ} &= \tilde{S}_\text{WZ} + \int d^2 x \, \beta \xi' \dot{\theta}, \label{S:WZ:U}
\end{align}
where the new constraint was implemented by an auxiliary field,
$\xi$, in Eq.~\eqref{S:cons}. In the last term~\eqref{S:WZ:U}, the
first one is the original WZ action~\eqref{S:WZ} and the second
one is due to the symmetry breaking of Eq.~\eqref{S:cons}.
Similarly to the previous case, the conserved Noether current
$j^\mu = J_{\rm N}^\mu + e \epsilon^{\mu\nu} A_\nu + \beta
\delta^\mu_1 \xi'$ is not gauge invariant. The gauge invariant
sector of it is the same as Eq.~(\ref{J:A}), which is not
conserved, $ \partial_\mu J_{\rm N}^\mu = - e \epsilon^{\mu\nu}
\partial_\mu A_\nu - \beta \xi''$.

Now, the equations of motion for $\beta \ne 0$ are obtained as
\begin{align}
  &\Box \phi + e \epsilon^{\mu\nu} \partial_\mu A_\nu - \beta (\lambda'' - \xi'') =
  0, \label{eom:dual:phi} \\
  &\lambda + \phi'' - eA'_0 - \dot\theta' = 0, \label{eom:dual:lambda}
  \\
  &\phi'' + eA'_0 + \dot\theta' = 0, \label{eom:dual:xi} \\
  &\dot\lambda' + \dot\xi' =
  0, \label{eom:dual:theta} \\
  &\partial_\mu F^{\mu\nu} = -e \left[ \epsilon^{\mu\nu} \partial_\mu \phi +
    \beta \epsilon^{1\nu} (\lambda' + \xi')
  \right]. \label{eom:dual:A}
\end{align}

For the explicit counting of degrees of freedom, we perform the
constraint analysis. Then, we get four primary constraints as
\begin{align}
  \Omega_1 &\equiv \Pi^0 \approx 0, \quad
  \Omega_2 \equiv \Pi_\lambda \approx 0, \label{Omega:1,2} \\
  \Omega_3 &\equiv \Pi_\xi \approx 0, \quad
  \Omega_4 \equiv \Pi_\theta -\beta (\lambda' +
  \xi') \approx 0, \label{Omega:3,4}
\end{align}
where $\Pi_\xi$ and $\Pi_\theta$ are the momenta of $\xi$ and
$\theta$, respectively.  By the use of the Legendre transformation,
the primary Hamiltonian can be obtained as
\begin{equation}
  \label{H:p}
  H_p = \int dx\, \left( \mathcal{H}_c + \sum_{i=1}^4 \eta_i \Omega_i \right),
\end{equation}
where the canonical Hamiltonian is
\begin{align}
  \mathcal{H}_c = &\frac12\left[ (\Pi_\phi + eA_1)^2 +
    \phi'^2 - \beta(\lambda^2 - 2 \lambda' \phi') \right] -
  e( \phi' + \beta \lambda') A_0 + \frac12 (\Pi^1)^2 + \Pi^1 A'_0 \notag \\
  &- \beta \xi' (\phi'+ eA_0),
\end{align}
and $\eta_i$'s are multiplier fields. Next, we choose the gauge fixing
condition as
\begin{equation}
  \label{Omega:5}
  \Omega_5 \equiv \theta \approx 0.
\end{equation}
Then, the time evolution of the four primary constraints and the gauge fixing
condition yield the following secondary constraints,
\begin{align}
  \Omega_6 &\equiv (\Pi^1)' + e[\phi' + \beta(\lambda' +
  \xi')] \approx 0, \label{Omega:6} \\
  \Omega_7 &\equiv \lambda+ \phi'' - eA'_0 \approx 0,
  \quad
  \Omega_8 \equiv \phi''+ e A'_0 \approx 0, \label{Omega:7,8}
\end{align}
and the Lagrange multipliers are completely fixed.

The Dirac bracket~\cite{Dirac:1964LQM} between two fields
$\mathcal{A}(x)$ and $\mathcal{B}(y)$ is also defined by
\begin{equation}
  \label{def:DB}
  \{\mathcal{A}(x), \mathcal{B}(y) \}_{\rm D} \equiv \{ \mathcal{A}(x),
  \mathcal{B}(y) \} - \sum_{i,j} \int dz dw \{ \mathcal{A}(x), \Omega_i(z) \}
  \Delta^{-1}_{ij} (z,w) \{ \Omega_j(w), \mathcal{B}(y) \},
\end{equation}
where $\Delta^{-1}_{ij}(x,y)$ is the inverse of $\Delta_{ij}
(x,y)$ defined by $\Delta_{ij} (x,y) = \{ \Omega_i(x), \Omega_j
(y) \}$.
The non-vanishing Dirac brackets can be calculated as
\begin{align}
  \{ A_0(x) , \Pi_\phi (y) \}_{\rm D} &= - \frac{1}{e} \delta'(x-y),
  \hspace{2.1cm}
  \{ A_1 (x), \xi (y) \}_{\rm D} =
  - \frac{1}{\beta e} \delta (x-y), \\
  \{ A_1(x), \Pi^1(y) \}_{\rm D} &=
  \delta (x-y), \hspace{2.5cm}
  \{ A_1(x), \Pi_\theta(y) \}_{\rm D} =
  \frac{1}{e} \delta'(x-y), \\
  \{ \phi(x), \Pi_\phi(y) \}_{\rm D} &= \delta(x-y), \hspace{2.7cm}
  \{ \lambda(x), \Pi_\phi(y) \}_{\rm D} =  -2 \delta''(x-y), \\
  \{ \xi(x), \Pi_\phi(y) \}_{\rm D} &= - \frac{1}{\beta} \delta (x-y)
  + 2 \delta''(x-y), \quad
  \{ \Pi_\phi(x), \Pi_\theta(y) \}_{\rm D} = - \delta'(x-y).
\end{align}
Note that they are well-defined for $\beta \ne 0$.

Now, the reduced Hamiltonian taking into account all constraint is
obtained as
\begin{align}
  H_{\rm red} = \frac12 \left[ (\Pi_\phi + eA_1)^2 + \left(\Pi^1 - \frac{1}{e}
  \phi'' \right)^2 + 3 \phi'^2 + \frac{4\beta e^2 -1}{e^2} \phi''^2
  \right],
\end{align}
which can be positive definite for $\beta \ge 1/(4e^2)$.  Note
that for $\beta \to 0$, the system does not go back to the reduced
Hamiltonian of the SM because the constraint system for $\beta\ne
0$ is different from that for $\beta=0$. Actually, there exist
more degrees of freedom compared to the case of the SM. This fact can
be easily seen by solving Eqs.~(\ref{eom:dual:phi})--
(\ref{eom:dual:A}),
\begin{equation}
  \label{eom:U:Fdual}
  \left( \Box + 4 \beta \partial_x^4 + e^2 \right)
  {}^*F = \mathcal{J}, \qquad \Box \mathcal{J} = 0,
\end{equation}
where the field $\mathcal{J}$ is related to the original fields by
$\mathcal{J} = \left( \Box - \partial_x^2 + 4 \beta \partial_x^4
\right) ({}^*F - e \phi)$.  Note that there are two modes: one is
a massless mode and the other is a massive mode with mass $e^2$
with the modified dispersion relation.  As a result, the degrees
of freedom is not conserved for $\beta \rightarrow 0$, which
implies that the SM can not be reproduced for the IR limit
in spite of the consistent formulation.

\section{Conclusion}
\label{sec:conclusion} In summary, we have presented two gauged
Lifshitz scalar field theory models. Actually, our criteria is
just whether the duality between the source current and the
Noether current is preserved or not, which was clearly satisfied
in the SM. For the case of the partial duality in
section~\ref{sec:gaugedLifshitz}, the SM limit is well-defined, so
that the massive single mode survives with the positivity of the
Hamiltonian. In this case, the characteristic parameter
``$\beta$'' of the higher-derivative in the Lifshitz scalar field
theory can increase the effective mass, where the lower bound is
$e^2$. This model is plausible in the sense that it is
continuously interpolated with the SM, but, unfortunately, the
duality relation is not complete. The second model in
section~\ref{sec:dual} seems to be nice in that the duality
relation is preserved, but it can not be smoothly connected with
the SM. To get the SM in the second case, we have to
begin with $\beta=0$.
The final result shows that there exist one massive and
one massless modes. It is rather close to the chiral
SM~\cite{Jackiw:1984zi, Rajaraman:1985qy, Girotti:1986ck,
Kim:1992ey,
  Banerjee:1993pj} from the view-point of degrees of
freedom. Actually, the additional massless degree of freedom is
due to the non-conservation of the source current, which spoils
the local gauge invariance and the more degree of freedom
survives.

So far, we have discussed the gauged Lifshitz scalar field
theories based on the duality relation. Strictly speaking, they
are not unique, so it will be interesting to find the other
Lifshitz scalar models to match the SM for the IR limit without
resort to the duality relation, and discuss some physical
applications. We hope this issue will be addressed elsewhere.

\begin{acknowledgments}
  M.~Eune was supported by National Research Foundation of Korea Grant
  funded by the Korean Government (Ministry of Education, Science and
  Technology) (NRF-2010-359-C00007).  W.~Kim was supported by the
  Basic Science Research Program through the National Research
  Foundation of Korea(NRF) funded by the Ministry of Education,
  Science and Technology(2010-0008359), and the Special Research Grant
  of Sogang University, 200911044.  E.\ J.\ Son was supported by the
  National Research Foundation of Korea(NRF) grant funded by the Korea
  government(MEST) through the Center for Quantum Spacetime(CQUeST) of
  Sogang University with grant number 2005-0049409.
\end{acknowledgments}



\end{document}